\documentclass[12pt]{iopart}
 \expandafter\let\csname equation*\endcsname\relax
 \expandafter\let\csname endequation*\endcsname\relax
\usepackage{amsmath}
\usepackage{cite}

\begin{document}

\title[Fidelity and purity of quantum electrical circuit states]
{Fidelity and purity of quantum electrical circuit states and
quantum tomograms}
\author{Olga V. Man'ko}
\address{P. N. Lebedev Physical Institute, Leninskii Prospect 53, Moscow
119991, Russia} \ead{omanko@sci.lebedev.ru}

\begin{abstract}
Review of the probability representation of quantum mechanics and
the symplectic tomography approach are presented. The examples of
Gaussian states of nanoelectric circuit, Josephson junction, and two
interacting high-quality resonant circuits are considered. The
Shannon entropy, quantum information, fidelity, and purity of
quantum states in the tomographic representation of quantum
mechanics are studied.
\end{abstract}


\vspace{2mm}

\section{Introduction}
Recently the theoretical aspects of the dynamical Casimir effect
were intensively studied (see,
e.g.~\cite{Dodonov,Dodonov1,DodonovSasha}. The experimental results
on the dynamical Casimir effect were obtained in
\cite{JohansonNature}. The important ingredient of this
investigation is the consideration of the effects related to the
behaviour of nonstationary quantum oscillators. The oscillators can
be realized by electric circuits or Josephson junctions with
time-varying
parameters~\cite{DMMPLPI1991,VVDOVMVIMJRLR1989,VVDOVMVIMJRLR1992,mestec1990,vimjrlr1991,ovmjkoreanphyssoc}.
The squeezing and other quantum properties of such devices were
discussed in
\cite{Zeilinger,Zeilinger1,Zeilinger2,DMMPLPI1991,VVDOVMVIMJRLR1989,mestec1990}.
The quantum states of the devices can be associated with tomographic
probability distributions (tomograms) considered for the
oscillators, e.g., in \cite{Ibort,Mancini}. Since the quantum
tomograms are the fair probability distributions, one can introduce
such characteristics of the quantum states as tomographic Shannon
entropy and information (see e.g.
\cite{ovmvimjrlr1997,mamjrlr2001,mavifoundphys}).

The aim of this work is to consider the application of the
tomographic approach to the problem of circuits and Josephson
junctions. We obtain an explicit form of the ground state tomogram
for two interacting high-quality resonant circuits as well as their
tomographic entropy and the Shannon information expressed in terms
of the quantum tomograms. For the Josephson junction modeled by a
resonant circuit, we obtain the tomogram of a ground-like Gaussian
state excited due to the time dependence of the critical current.

The paper is organized as follows.

In the first section, we discuss a quantum resonant circuit in the
probability representation of quantum mechanics. In the second
section, we consider the Josephson junction modeled by resonant
circuit with time-dependent parameters, discuss the quantum tomogram
of its Gaussian state and an analog of the nonstationary Casimir
effect in this system. In the third section, we discuss two
high-quality interacting resonant circuits and obtain the symplectic
tomogram of their Gaussian state.

\section{Quantum resonant circuit}
Following \cite{DMMPLPI1991,Vaxjo2011}, we present a short review of
high-quality resonant circuits with inductance $L$ and capacity $C$
in the tomographic-probability representation of quantum mechanics.
The resonant circuit must be considered as the quantum one, if the
energy of thermal fluctuations is smaller then the energy of
vibration quanta $ \hbar\omega>k T,$ where $T$ is the temperature
and $k$ is the Boltzmann constant. For example, if the wavelength is
$\sim$1~cm and the temperature is lower than 1.4~K, the resonant
circuit must be considered as a quantum object. The Hamiltonian of
the resonant circuit is of the form (see, e.g., \cite{DMMPLPI1991})
\begin{equation}\label{hamiltonianreonantscircuit}
\hat H=\frac{1}{2}\left(\frac{\hat Q^2}{C}+L\hat I^2\right),
\end{equation}
where $\hat Q$ is an operator of charge on the plates of a capacitor
and $\hat I$ is an operator of the current in the circuit. The
commutation relation for the operators of charge and current is of
the form
\begin{equation}\label{commutation relation}
[\hat I,\hat Q]=i\hbar/L.
\end{equation}
For an operator of voltage on the plates of the capacitor $\hat
V=\hat Q / C$, one has the commutation relation with operator of
current $[\hat I,\hat V]=i\hbar\omega^2,$ where $\omega=(L
C)^{-1/2}$ is the plasma frequency.

In \cite{Mancini}, the probability representation of quantum
mechanics was introduced and applied for considering the photon
states. It was shown that the quantum state can be determined by the
symplectic tomogram, which is expressed through the Wigner function
of the state. Applying usual symplectic-tomography-scheme
procedure~\cite{Mancini} to the problem of quantum circuit and
following \cite{Vaxjo2011}, we consider the general linear
combination of the current and voltage
\begin{equation}\label{observable}
\hat J=\frac{\mu\hat I}{i_0}+\frac{\nu\hat V}{u_0}\,,
\end{equation}
where $\hat I$ is the operator of current, $\hat V$ is the operator
of voltage, $\mu$ and $\nu$ are real numbers, which label the
reference frame in the current--voltage space $(I,V)$ of the system,
$i_0=\sqrt{\hbar\omega/L}$ and $u_0=\sqrt{\hbar/\omega LC^2}$ are
the amplitudes of vacuum fluctuations of current and voltage. Thus,
in our approach, the current and voltage are considered as analogs
of the photon quadrature components. The linear
combination~(\ref{observable}) is an  analog of the photon homodyne
quadrature. We assume that the current and voltage are dimensionless
ones. The state of the resonant circuit can be determined by the
symplectic tomogram, which is a function of the observable $J$ and
additional variables $\mu$ and $\nu$ and is expressed through the
Wigner function~\cite{Wig32} as follows~\cite{Vaxjo2011}:
\begin{equation}\label{radontrdirect}
 w\left(J,\mu,\nu\right )=\frac{1}{2\pi}\int
W(I,V)\delta(J-\mu I-\nu V)\,dI\,dV.
\end{equation}
The symplectic tomogram is normalized and nonnegative and has all
the properties of the standard probability distribution function.
For example, the tomogram of the resonant-circuit ground state is
given by normal distribution of the variable $J$ with the dispersion
determined by the parameters $\mu$ and $\nu$,
\[
w_0(J,\mu,\nu)=\frac{1}{\sqrt{\pi(\mu^2+\nu^2)}}
\exp\left(-\frac{J^2}{\mu^2+\nu^2}\right).
\]
The information contained in tomogram $w\left(J,\mu,\nu \right)$ is
overcomplete. For $\mu=\cos\theta$ and $\nu=\sin\theta$, where
$\theta$ is the rotation angle of the axis in the $(I,V)$ plane, the
symplectic tomogram is analogous to the optical tomogram introduced
in \cite{Bertran} and measured in \cite{Raymer} where the experiment
on reconstructing the photon Wigner function was performed.

The Wigner function can be reconstructed from the symplectic
tomogram, in view of the inverse Radon transform,
\begin{equation}\label{radontrinverse} W(I,V)=\frac
{1}{2\pi }\int w\left (J,\mu ,\nu \right) \exp \left [-i\left (\mu
I+\nu V-J\right )\right ] d\mu \,d\nu \,dJ.
\end{equation}
The density operator also can be reconstructed from
the symplectic tomogram
\[
\hat\rho=\frac{1}{2\pi}\int w(J,\mu,\nu)e^{i(J-\mu\hat I-\nu\hat
V)}\,dJ \,d\mu\, d\nu.
\]
Here we adopted the formula employed in quantum
optics~\cite{Ariano}. Due to the fact that the symplectic tomogram
is the probability distribution function, one can introduce entropy
and information following the Shannon prescription. The entropies
associated with different types of tomograms were discussed in
\cite{ovmvimjrlr1997,mamjrlr2001,NicolaFedeleMAMVIMEurPhysJ2003,OVMVIMJRLR2004,OVMJRLR2007,OVMNVTSPIE2006}
and were named the probability-representation entropies. The
tomographic entropy, associated with the resonant-circuit state and
determined by the symplectic tomogram, reads
\begin{equation}\label{entropy}
S(\mu,\nu)=-\int w(J,\mu,\nu)\ln w(J,\mu,\nu)\, dJ.
\end{equation}
The tomographic entropy is a function of additional parameters which
label a reference frame in the $(I,V)$ space.

Such known quantities as the fidelity and purity also can be
calculated for the resonant-circuit state, in view of the state
symplectic tomogram following~\cite{Fortschrih}. Using the
connection of the density operator of the resonant-circuit state and
its symplectic tomogram, we obtain for the fidelity
\begin{eqnarray}\label{fidelity}
\mbox{\cal F}=\mbox{Tr}\left(\hat\rho_1\hat\rho_2\right)
&=&\frac{1}{2\pi}\int w_1(J_1,\mu_1,\nu_1)w_2(J_2,-\mu_1,-\nu_1)\nonumber\\
&&\times\exp\left(i(J_1+J_2)\right)
dJ_1\,dJ_2\,d\mu_1\,d\mu_2\,d\nu_1\,d\nu_2,
\end{eqnarray}
where $\hat\rho_1$ and $\hat\rho_2$ are the density operators and
$w_1(J_1,\mu_1,\nu_1)$ and $w_2(J_2,-\mu_1,-\nu_1)$ are the
tomograms associated with the first and second states of the
resonant circuit, respectively.

The purity of the resonant-circuit state, which we denote
$\mbox{\cal P}$, reads
\begin{equation}\label{purity}
\mbox{\cal P}=\mbox{Tr}\left(\hat\rho^2\right)=\frac{1}{2\pi}\int
w(J_1,\mu_1,\nu_1)w(J_2,-\mu_1,-\nu_1) \exp\left[i(J_1+J_2)\right]
dJ_1\,dJ_2\,d\mu_1\,d\mu_2\,d\nu_1\,d\nu_2.
\end{equation}
The formula for purity written in terms of measurable optical
tomograms (using the expression of symplectic tomogram in terms of
the optical tomogram)~\cite{Fortschrih,vimmarmophysscr} was used in
the experiments~\cite{bellini} on homodyne detection of some photon
states.

For the fidelity and purity of quantum states of real physical
resonant circuit, we have the inequalities
\begin{eqnarray}
\fl 0\leq\frac{1}{2\pi}\int
w_1(J_1,\mu_1,\nu_1)w_2(J_2,-\mu_1,-\nu_1)
\exp\left(i(J_1+J_2)\right)dJ_1\,d\,J_2\,d\mu_1\,d\mu_2\,d\nu_1\,d\nu_2
\leq 1,\label{enequalityfidelity}\\
\fl 0\leq\frac{1}{2\pi}\int w(J_1,\mu_1,\nu_1)w(J_2,-\mu_1,-\nu_1)
\exp\left(i(J_1+J_2)\right)dJ_1\,dJ_2\,d\mu_1\,d\mu_2\,d\nu_1\,d\nu_2
\leq 1.\label{inequalitypurity}\end{eqnarray} The tomograms
associated with the states of a real physical resonant circuit must
satisfy the inequality, which is the nonnegativity condition of the
density operator,
\begin{equation}\label{uinequalitydensityoperatornonnegativity}
\int w(J,\mu,\nu)\exp\left(i(J\hat1-\mu \hat I-\nu\hat Q)\right) dJ\, d\mu\,d\nu\geq 0.
\end{equation}
The obtained expressions for fidelities and purities are given in
terms of measurable tomographic probability distributions. They can
be applied for checking the quantumness of the states in experiments
which are analogs of the homodyne detection of photon states. The
Josephson-junction realization of circuit QED can be used to study
quantum properties of the circuits.

\section{Parametric Josephson junction}
As an application of our model we discussed above, we consider such
object as the Josephson junction. The Josephson junction at zero
temperature is described by the Hamiltonian~\cite{Anderson,Likharev}
\begin{equation}\label{hamiltonianjosephsonjunction}
\hat H=\frac{\hat Q^2}{2C}+\frac{\hbar
I_c(t)}{2e}\left(1-\cos\hat\phi\right)-\frac{\hbar I_k(t)\hat\phi}{2
e},
\end{equation}
where $C$ is the capacitance of the junction, $I_c$ is the critical
current, $\hat Q$ is the charge operator, $\hat\phi$ is the phase
operator, $I_k(t)$ is the external classical current, $e$ is the
electron charge, and $\hbar$ is the Planck constant.

Now we consider the Josephson junction in the domain of small
phases. Also we assume that the shot noise is smaller than the
quantum fluctuations, $\delta Q\gg 2e.$ This means that we have the
following condition for the critical current and capacity of the
Josephson junction:
\begin{equation}\label{condition}
I_c C\gg \frac{32e^3}{\hbar}\sim 10^{-21}\Phi A.
\end{equation}
Under the condition~(\ref{condition}), the term $1-\cos\hat\phi$ in
the Hamiltonian~(\ref{hamiltonianjosephsonjunction}) can be replaced
by the quadratic expression $\hat\phi^2/2$. So instead of the
Hamiltonian~(\ref{hamiltonianjosephsonjunction}) we obtain the
Hamiltonian of a driven resonance circuit
\begin{equation}\label{hamiltonianquadraticjosephson}
\hat H=\frac{\hat Q^2}{2C}+\frac{\hbar I_c(t)\hat\phi^2}{2e}-\frac{\hbar I_k(t)\hat\phi}{2 e}.
\end{equation}
The commutation relation for the charge and phase operators reads
\begin{equation}
[\hat\phi,\hat Q]=2i e,\end{equation} that provides the following
commutation relation for the current $\hat I=-I_c\hat\phi$ and
voltage $\hat V=\hat Q/C$ operators:
\begin{equation} [\hat I,\hat V]=i\hbar\omega^2,
\end{equation}
where
\begin{equation}\label{plasma}
\omega=\left({2e I_c}/{\hbar C}\right)^{1/2}
\end{equation}
is the plasma frequency of the junction. If the parameters (critical
current or capacity) of the junction are dependent on time, then the
plasma frequency is also the function of time, and so the Josephson
junction is acted by a parametric excitation. Below we assume such
the units that provide all the variables like the current and
voltage to be dimensionless. The Josephson-junction Gaussian state
can be determined by the tomogram
\begin{equation}\label{josephsonground statetomogramm}
w(J,\mu,\nu,t
)=\frac{1}{\sqrt{2\pi\sigma_J(t)}}\exp\left(-\frac{(J-\bar{J})^2}{2\sigma_J(t)}\right),\end{equation}
where the dispersion of the observable is expressed through the
dispersions of the current and voltage as follows:
\begin{equation}\sigma_J(t)=\mu^2\sigma_{I^2}(t)+\nu^2\sigma_{V^2}(t)+2\mu\nu\sigma_{I V}(t),\end{equation}
and the dispersions of the current and voltage are
\begin{equation}\label{dis}
\sigma_{I^2}(t)= {|\epsilon(t)|^2}/{2},\quad
\sigma_{V^2}(t)={|\dot\epsilon(t)|^2}/{2},\quad \sigma_{I
V}=\sqrt{\sigma_{I^2}\sigma_{V^2}-{1}/{4}}\,.\end{equation} The
function $\epsilon(t)$ in (\ref{dis}) satisfies the equation
\begin{equation}\label{epsilon}
\ddot\epsilon(t)+\omega^2(t)\epsilon(t)=0,
\end{equation}
with additional condition
\begin{equation}
\dot\epsilon\epsilon^\ast-\dot\epsilon^\ast\epsilon=2\,i,
\end{equation}
where the time-dependent frequency in Eq.~(\ref{epsilon}) is given
by (\ref{plasma}) with the time-dependent critical current. The mean
values of observable $J$ is
\[\bar J=-\sqrt 2\left[\mu(\mbox{Re}\left(\delta\epsilon^*\right)+\nu\mbox{Re}\left(\delta\dot\epsilon^*\right)\right],\]
where function $\delta(t)$ is
\[\delta(t)=-\frac{i}{\sqrt2}\int_0^t I_k(\tau)\epsilon(\tau)\, d\tau.
\]
In such a system, an analog of the well-known Casimir effect can
exist. The Casimir force is an attraction force which exists between
two noncharged plates without photons between them. The Casimir
force exists due to the dependence of the electromagnetic-field
vacuum energy on the geometry of the system. The vacuum energy
depends on the parameters of the system. This means, that the forces
exist which try to minimize the vacuum energy. In the parametric
case where the system has time-dependent parameters, the
nonstationary Casimir effect appears. The energy of external
mechanical source goes to reforming the vacuum energy, and the
electromagnetic radiation appears \cite{DOD}. In
\cite{VVDOVMVIMJRLR1989,VVDOVMVIMJRLR1992,mestec1990,vimjrlr1991,ovmjkoreanphyssoc},
it was suggested to use the parametric Josephson junction for
obtaining an analog of the nonstationary Casimir effect. Due to
nonstationary Casimir effect, there exists a possibility to obtain
electric oscillations without connecting the junction with external
electromagnetic sources due to changing the inductance or capacity
of the junction. Applying the external energy, we change the
critical current or capacity of the junction and obtain the energy
of electric oscillations. So, the Josephson junction with varying
parameters can be used as a quantum generator of current, as was
suggested in
\cite{VVDOVMVIMJRLR1989,VVDOVMVIMJRLR1992,mestec1990,vimjrlr1991,ovmjkoreanphyssoc}.

Nowadays, the discussions of analogous effects are known as the
circuit quantum electrodynamics. These effects were studied, e.g.,
in \cite{Zeilinger,Zeilinger1,Zeilinger2,Mallet,Walraff}.

\section{Two interacting resonant circuits}
In this section, we obtain an explicit form of the two-mode
symplectic tomogram of the ground quantum state for two interacting
resonant circuits. We also obtain new tomographic characteristics
for this state such as the tomographic entropy and information. We
consider two high-quality interacting resonant circuits with
inductances $L_n$ and capacities $C_n$. The Hamiltonian of the
system is of the form
\begin{equation}\label{hamiltoniantwocircuits}
\hat H=\frac{L \hat I_1^2}{2}+\frac{\hat Q_1^2}{2C}+
\frac{L \hat I_1^2}{2}+\frac{\hat Q_1^2}{2C}+L_{12}\hat I_1\hat I_2,
\end{equation}
where $\hat Q_1$ and $\hat Q_2$ are the operator of charge on plates
of the capacitor in the first and second circuits, $\hat I_1$ and
$\hat I_2$ are operators of current in the first and second
circuits, respectively, and $L_{12}$ is the selfinductance. The
commutation relations for operators of charges and currents read
\begin{equation}\label{commutationrelation2rc}
[\hat I_1,\hat Q_1]=i\hbar/L,\qquad[\hat I_2,\hat
Q_2]=i\hbar/L,\qquad [\hat I_1,\hat Q_2]=[\hat I_2,\hat Q_1]=0.
\end{equation}
For operators of voltage on the plates of capacitors $\hat V_1=\hat
Q_1/C$ and $\hat V_2=\hat Q_2/C$, one has the commutation relations
with operators of currents $[\hat I_1,\hat V_1]=[\hat I_2,\hat
V_2]=i\hbar\omega^2,$ where $\omega$ is the plasma frequency.
Introducing the new variables
\[\hat I_k=\left({\hat I_1+\hat I_2}\right)/{\sqrt2},\qquad
\hat I_s=\left({\hat I_1-\hat I_2}\right)/{\sqrt2},
\]
we obtain the Hamiltonian (\ref{hamiltoniantwocircuits}) in the form
\begin{equation}\label{hamiltoniantwocircuitsin}
\hat H=\frac{L_k \hat I_k^2}{2}+\frac{\hat Q_k^2}{2C}+\frac{L_s \hat
I_s^2}{2}+\frac{\hat Q_s^2}{2C}\,,
\end{equation}
where $L_k=L+L_{12}$ and $ L_s=L-L_{12}$. The frequencies of
oscillations of two noninteracting resonant circuits with the
Hamiltonian~(\ref{hamiltoniantwocircuitsin}) are $
w_s=\omega\sqrt{L/L_s}$ and $w_k=\omega\sqrt{L/L_k},$ and new
amplitudes of the vacuum fluctuations read
\[i_k=\sqrt{\hbar \omega/L_k},\qquad q_k=\sqrt{\hbar/\omega L_k},\qquad
i_s=\sqrt{\hbar \omega/L},\qquad q_s=\sqrt{\hbar/\omega L_s}.\]

The system of two interacting high-quality resonant circuits can
also be considered within the framework of symplectic-tomography
scheme~\cite{Mancini}. Below in this section and Appendix, we assume
the current, charge, and voltage to be dimensionless. We introduce
the general linear combinations of currents and charges as follows:
\begin{equation}\label{observable2c}
\hat J_1=\mu_1\hat I_1+\nu_1\hat Q_1,\quad \hat J_2=\mu_2\hat I_2+\nu_2\hat Q_2,
\end{equation}
where $\mu_1$, $\mu_2$, $\nu_1$, and $\nu_2$ are real numbers, which
label a reference frame in the currents--charge space $(I_1,I_2,
Q_1, Q_2)$ of the system.

The state of the system of two resonant circuits can also be
determined by symplectic tomogram, which is the function of
observables $J_1$ and $J_2$ and additional variables $\mu_1$,
$\mu_2$, $\nu_1$, and $\nu_2$. For example, the symplectic tomogram
of the system Gaussian state is
\begin{equation}\label{tom2c}
w(J_1,J_2,\mu_1,\mu_2,\nu_1,\nu_2)=\frac{1}{2\pi\sqrt{\det
\sigma(t)}} \exp\left(-\frac{1}{2}\bf{J}\sigma^{-1}(t)\bf{J}\right),
\end{equation}
where components of the vector $\bf{J}$ are observables $J_1$ and
$J_2$ and the dispersion matrix $\sigma(t)$
\[\sigma(t)=\Big(\begin{array}{cc}
\sigma_{J_1J_1}&\sigma_{J_1J_2} \\
  \sigma_{J_1J_2}&\sigma_{J_2J_2}
\end{array}\Big)
\]
are expressed through the dispersion of current and charge in the form
\begin{eqnarray}\label{dispersionobservable2c}
&&\sigma_{J_1J_1}=\mu_1\sigma_{I_1^2}(t)
+\nu_1\sigma_{Q_1^2}(t)+2\mu_1\nu_1\sigma_{I_1Q_1}(t),\nonumber\\
&&\sigma_{J_2J_2}=\mu_2\sigma_{I_2^2}(t)+\nu_2\sigma_{Q_2^2}(t)
+2\mu_2\nu_2\sigma_{I_2Q_2}(t),\label{dispersionobservable2c}\\
&&\sigma_{J_1J_2}=\mu_1\nu_2\sigma_{I_1Q_2}(t)+\mu_2\nu_1\sigma_{I_2Q_1}(t)
+\mu_1\mu_2\sigma_{I_1I_2}(t)+\nu_1\nu_2\sigma_{Q_1Q_2}(t).\nonumber
\end{eqnarray}
Explicit expressions for the current--charge dispersion-matrix
elements are given in Appendix by
formulae~(\ref{dispersioncharge2c})--(\ref{koef}).
%(\ref{covariances2c}).

Due to the fact that the symplectic tomogram is the probability
distribution function, following the Shannon prescription, we can
associate the tomographic entropy~\cite{ovmvimjrlr1997} with the
state of two interacting circuits described by the tomogram in the
probability representation of quantum mechanics as follows:
\begin{equation}\label{entropy2c}
S(\mu_1,\mu_2,\nu_1,\nu_2)=- \int
w(J_1,J_2,\mu_1,\mu_2,\nu_1,\nu_2)\ln
w(J_1,J_2,\mu_1,\mu_2,\nu_1,\nu_2)\,dJ_1\, dJ_2.
\end{equation}
Also we can introduce the tomographic information associates with
the system state
\begin{equation}\label{inform}
{\cal I}=\int
w(J_1,J_2,\mu_1,\mu_2,\nu_1,\nu_2)\ln\Big[\frac{w(J_1,J_2,\mu_1,\mu_2,\nu_1,\nu_2)
}{ w_1(J_1,\mu_1,\nu_1) w_2(J_2,\mu_2,\nu_2) }\Big]\, dJ_1\,d J_2,
\end{equation}
where \begin{eqnarray*} w_1(J_1,\mu_1,\nu_1,)=\int
w(J_1,J_2,\mu_1,\mu_2,\nu_1,\nu_2)\,dJ_2,\\
w_2(J_2,\mu_2,\nu_2)=\int
w(J_1,J_2,\mu_1,\mu_2,\nu_1,\nu_2)\,dJ_1.\end{eqnarray*}

Thus, we got two tomographic characteristics of two circuit states
describing the degree of quantum correlations in the system state,
such as entropy and information.

\section{Conclusions}
Concluding, we point out the main results of our study.

We reviewed the notion of quantum state in the symplectic tomography
approach on the examples of two interacting high-quality circuits
and Josephson junctions and discussed an analog of the nonstationary
Casimir effect in the Josephson junction. We presented the entropy,
information, fidelity, and purity of the circuit states in the
tomographic-probability representation. For the Gaussian state of
the resonant circuit, we introduced the tomographic entropy and
expressed the entropy in terms of the symplectic tomogram of the
resonant-circuit state. For two different quantum resonant circuit
states, we obtained the expression for fidelity in terms of
symplectic tomograms of these states. Particulary, we got the
expression for purity of the resonant circuit state in terms of
symplectic tomograms. We got inequalities which determine the
quantumness of the circuit states. For the system of two
high-quality interacting resonant circuits, we introduced the
tomographic entropy and tomographic information.

\subsection*{Acknowledgments}
The study was supported by the Russian Foundation for Basic Research
under Project No.~10-02-00312. The author is grateful to the
Organizers of the 19th Central European Workshop on Quantum Optics
(Sinaia, Romania, 2--6 July 2012) and especially Prof. Aurelian Isar
for invitation and kind hospitality.

\section*{Appendix}
Assuming dimensionless variables $\omega_k$ and $\omega_s$ and
$\omega=1$, we calculate matrix elements of the dispersion matrix
\begin{eqnarray}%\label{dispersioncharge2c}
\fl\sigma_{Q_1^2}(t)=\frac{1}{4}
\big[c_+^2\sigma_{Q_1^2}(0)+c_-^2\sigma_{Q_2^2}(0)+k_{+}^2\sigma_{I_1^2}(0)+
k_-^2\sigma_{I_2^2}(0)+2\big(k_+k_-\sigma_{I_1I_2}(0)+c_+k_+\sigma_{I_1Q_1}(0)\nonumber\\
\fl+c_+c_-\sigma_{Q_1Q_2}(0)+k_-c_-\sigma_{I_2Q_2}(0)+k_-c_-\sigma_{Q_1I_2}(0)+k_+c_-\sigma_{I_1Q_2}(0)\big)\big],\nonumber\\
\fl\sigma_{Q_2^2}(t)=\frac{1}{4}
\big[c_-^2\sigma_{Q_1^2}(0)+c_+^2\sigma_{Q_2^2}(0)+k_{-}^2\sigma_{I_1^2}(0)+
k_+^2\sigma_{I_2^2}(0)+2\big(k_+k_-\sigma_{I_1I_2}(0)+c_-k_-\sigma_{I_1Q_1}(0)\nonumber\\
\fl+c_+c_-\sigma_{Q_1Q_2}(0)+k_-c_+\sigma_{I_1Q_2}(0)+k_+c_-\sigma_{Q_1I_2}(0)
-k_+c_+\sigma_{I_2Q_2}(0)\big)\big],\label{dispersioncharge2c}\\
\fl\sigma_{Q_1Q_2}(t)=\frac{1}{4}
\big[c_-c_+\big(\sigma_{Q_1^2}(0)+\sigma_{Q_2^2}(0)\big)
+k_{-}k_+\big(\sigma_{I_1^2}(0)+\sigma_{I_2^2}(0)\big)+(k_+^2+k_-^2)\sigma_{I_1I_2}(0)\nonumber\\
\fl+(c_-k_++k_-c_+)\big(\sigma_{I_1Q_1}(0)+\sigma_{I_2Q_2}(0)\big)
+(c_+^2+c_-^2)\sigma_{Q_1Q_2}(0)\nonumber\\
\fl
+(k_-c_-+k_+c_+)\big(\sigma_{I_1Q_2}(0)+\sigma_{Q_1I_2}(0)\big)\big].\nonumber
\end{eqnarray}
The dispersions of currents after interaction read
\begin{eqnarray}%\label{dispersioncurrent2c}
\fl\sigma_{I_1^2}(t)=\frac{1}{4}
\big[s_+^2\sigma_{Q_1^2}(0)+s_-^2\sigma_{Q_2^2}(0)+c_{+}^2\sigma_{I_1^2}(0)+
c_-^2\sigma_{I_2^2}(0)+2\big(c_+c_-\sigma_{I_1I_2}(0)
-c_+s_+\sigma_{I_1Q_1}(0)\nonumber\\
\fl+s_+s_-\sigma_{Q_1Q_2}(0)-s_-c_-\sigma_{I_2Q_2}(0)-s_+c_-\sigma_{Q_1I_2}(0)
-c_+s_-\sigma_{I_1Q_2}(0)\big)\big],\nonumber\\
\fl\sigma_{I_2^2}(t)=\frac{1}{4}
\big[s_-^2\sigma_{Q_1^2}(0)+s_+^2\sigma_{Q_2^2}(0)+c_{-}^2\sigma_{I_1^2}(0)+
c_+^2\sigma_{I_2^2}(0)+2\big(c_+c_-\sigma_{I_1I_2}(0)-c_-s_-\sigma_{I_1Q_1}(0)\nonumber\\
\fl+s_+s_-\sigma_{Q_1Q_2}(0)-c_-s_+\sigma_{I_1Q_2}(0)-s_-c_+\sigma_{Q_1I_2}(0)
-s_+c_+\sigma_{I_2Q_2}(0)\big)\big],\label{dispersioncurrent2c}\\
\fl\sigma_{I_1I_2}(t)=\frac{1}{4}
\big[s_-s_+\big(\sigma_{Q_1^2}(0)+\sigma_{Q_2^2}(0)\big)
+c_{-}c_+\big(\sigma_{I_1^2}(0)+\sigma_{I_2^2}(0)\big)+(c_+^2+c_-^2)\sigma_{I_1I_2}(0)\nonumber\\
\fl-(c_-s_++s_-c_+)\big(\sigma_{I_1Q_1}(0)+\sigma_{I_2Q_2}(0)\big)
+(s_+^2+s_-^2)\sigma_{Q_1Q_2}(0)\nonumber\\
\fl
+(s_-c_-+s_+c_+)\big(\sigma_{I_1Q_2}(0)+\sigma_{Q_1I_2}(0)\big)\big].\nonumber
\end{eqnarray}
The covariances are
\begin{eqnarray}%\label{covariances2c}
\fl\sigma_{I_1Q_1}(t)=\frac{1}{4}
\big[-c_+s_+\sigma_{Q_1^2}(0)-s_-c_-\sigma_{Q_2^2}(0)+c_+k_
+\sigma_{I_1^2}(0)+c_-k_-\sigma_{I_2^2}(0)\nonumber\\
\fl-(c_-s_++s_-c_+)\sigma_{Q_1Q_2}(0)+(c_-k_++c_+k_-)\sigma_{I_1I_2}(0)+(c_+^2-k_+s_+)\sigma_{I_1Q_1}(0)\nonumber\\
\fl+(c_-^2-s_-k_-)\sigma_{I_2Q_2}(0)
+(c_+c_--k_-s_+)\sigma_{Q_1I_2}(0)+(c_+c_--s_-k_+)\sigma_{I_1Q_2}(0)\big],\nonumber\\
\fl\sigma_{I_2Q_2}(t)=\frac{1}{4}
\big[-c_-s_-\sigma_{Q_1^2}(0)-s_+c_+\sigma_{Q_2^2}(0)+c_-k_-\sigma_{I_1^2}(0)+c_+k_+\sigma_{I_2^2}(0)\nonumber\\
\fl-(c_+s_-+s_+c_-)\sigma_{Q_1Q_2}(0)
+(c_-k_++c_+k_-)\sigma_{I_1I_2}(0)+(c_-^2-k_-s_-)\sigma_{I_1Q_1}(0)\nonumber\\
\fl
+(c_+^2-s_+k_+)\sigma_{I_2Q_2}(0)+(c_+c_--k_+s_-)\sigma_{Q_1I_2}(0)+(c_+c_--s_+k_-)\sigma_{I_1Q_2}(0)\big],\nonumber\\[-2mm]
\label{covariances2c}\\[-2mm]
\fl\sigma_{I_2Q_1}(t)=\frac{1}{4} \big[-
c_+s_-\sigma_{Q_1^2}(0)-s_+c_-\sigma_{Q_2^2}(0)
+c_-k_+\sigma_{I_1^2}(0)+c_+k_-\sigma_{I_2^2}(0)\nonumber\\
\fl
-(c_+s_++s_-c_-)\sigma_{Q_1Q_2}(0)+(c_+k_++c_-k_-)\sigma_{I_1I_2}(0)+(c_+c_--k_+s_-)\sigma_{I_1Q_1}(0)\nonumber\\
\fl +(c_+c_--s_+k_-)\sigma_{I_2Q_2}(0)
+(c_{+}^2-k_-s_-)\sigma_{Q_1I_2}(0)
+(c_{-}^2-s_+k_+)\sigma_{I_1Q_2}(0)\big],\nonumber\\
\fl\sigma_{I_1Q_2}(t)=\frac{1}{4} \big[-
c_-s_+\sigma_{Q_1^2}(0)-s_-c_+\sigma_{Q_2^2}(0)
+c_+k_-\sigma_{I_1^2}(0)+c_-k_+\sigma_{I_2^2}(0)\nonumber\\
\fl -(c_+s_++s_-c_-)\sigma_{Q_1Q_2}(0)
+(c_+k_++c_-k_-)\sigma_{I_1I_2}(0)+(c_+c_--k_-s_+)\sigma_{I_1Q_1}(0)\nonumber\\
\fl +(c_+c_--s_-k_+)\sigma_{I_2Q_2}(0)
+(c_{-}^2-k_+s_+)\sigma_{Q_1I_2}(0)+(c_{+}^2-s_-k_-)\sigma_{I_1Q_2}(0)\big],\nonumber
\end{eqnarray}
where
\begin{eqnarray}\label{koef}
\fl c_\pm=\cos\omega_k t\pm\cos\omega_s t,\qquad
k_\pm=\frac{\sin\omega_k t}{\omega_k}\pm \frac{\sin\omega_s
t}{\omega_s}\,,\qquad s_\pm=\sin\omega_k t\omega_k\pm\sin\omega_s
t\omega_s.\nonumber\\
\end{eqnarray}

\end{document}